\documentclass[12pt]{article}
\pdfoutput=1

\usepackage{epsfig,a4wide,amsmath,amssymb,amsfonts,mathrsfs}
\usepackage{mathtools}
\usepackage{color,yfonts}
\usepackage{graphicx}
\usepackage{subcaption}
\usepackage[utf8]{inputenc}
\usepackage{siunitx}
\usepackage{rotating}

\numberwithin{equation}{section}

\newcommand{\be}{\begin{equation}}

\newcommand{\ee}{\end{equation}}
\numberwithin{equation}{section}
\newcommand{\mytitlefont}{\fontseries{mx}\selectfont}
\DeclareMathAlphabet{\titlemath}{OT1}{cmr}{mx}{n}

\begin{document}

\begin{titlepage}

\begin{center}

~\\[2cm]

{\fontsize{20pt}{0pt} \mytitlefont  Gravitational Corner Conditions in Holography }

~\\[0.5cm]

{\fontsize{14pt}{0pt} Gary~T.~Horowitz and Diandian~Wang}

~\\[0.1cm]

\it{   Department of Physics, University of California, Santa Barbara, CA 93106}

~\\[0.05cm]

\end{center}

  \vspace{60pt}

\noindent
Contrary to popular belief,  asymptotically anti-de Sitter solutions of gravitational theories cannot be obtained by taking initial data (satisfying the constraints) on  a spacelike surface, and choosing an arbitrary conformal metric on the timelike boundary at infinity.  There are an infinite number of corner conditions that also must be satisfied where the initial data surface hits the boundary. These are well known to mathematical relativists, but to make them more widely known we give a simple explanation of why these conditions exist and discuss some of their consequences.  An example is given which illustrates their power. Some implications for holography are also mentioned.

\vfill

    \noindent

  \end{titlepage}

   \newpage

\baselineskip=16pt
\section{Introduction}

It appears to be widely believed that one can obtain smooth asymptotically anti-de Sitter (AdS) solutions of general relativity with a negative cosmological constant  by taking $C^\infty$ initial data satisfying the constraints on a spacelike surface, and choosing an arbitrary $C^\infty$ conformal metric on the timelike boundary at infinity. (If matter fields are present, one expects to be able to freely choose suitable boundary data for them as well as initial data.)  However, this is incorrect. Mathematical relativists have shown that there are an infinite number of conditions that must be satisfied at the corner where the initial data surface hits the asymptotic boundary \cite{Friedrich:1995vb,Enciso:2014lwa,Carranza:2018wkp}.\footnote{If the solution is not required to be smooth,  there are only a finite number of conditions \cite{Enciso:2014lwa}. This is discussed in Sec. 3.} If we label the initial data surface $t=0$, these conditions determine {\it all} time derivatives of the boundary conformal metric (and boundary data for the matter fields) at $t=0$. If the boundary data is analytic, it is completely determined by the initial data. There is still freedom in choosing boundary data, but it  cannot be analytic.

In this note we explain why these conditions exist, discuss some consequences, and give a simple example to illustrate them. To keep the discussion as simple as possible, we consider the case without matter, and focus mostly on four spacetime dimensions.
The need for an infinite number of corner conditions can already be seen in the simple case of the two dimensional wave equation on a half line. If $u=x-t$ and $v=x+t$ are null coordinates on Minkowski spacetime, and we are given initial data $\phi(x,t=0) = f(x)$ and $\dot \phi (x,t=0) = 0$ for $x<0$,  then the solution in the domain of dependence is simply $\phi(u,v) = [f(u) + f(v)]/2 $. The value of the field on the boundary $\phi(x=0,t)$ for $t>0$, depends on $f(v)$ for $v>0$ which is arbitrary except for the fact that $f(v)$ needs to be smooth at $v=0$. This means that all time derivatives of $\phi$ at ($x=0, t=0$) are determined in terms of space derivatives at this point.
  
We can see a similar effect in general relativity for asymptotically AdS solutions as follows. Consider smooth initial data on a complete spacelike surface. This consists of the spatial metric $g_{ij}$ and extrinsic curvature $k_{ij}$  satisfying the  constraints 
\be\label{constraint}
{\cal R} -k_{ij}k^{ij} + k^2 = 2\Lambda, \qquad D_j(k^{ij} - kg^{ij}) = 0,
\ee
where ${\cal R}$ is the three dimensional scalar curvature, $D_i$ is the three dimensional covariant derivative, and $k$ is the trace of $k_{ij}$.  Assuming\footnote{This simple choice is sufficient for our purposes of computing time derivatives, but it is not ideal for actual evolution since the equations are not strongly hyperbolic.}   $g_{tt} = -N^2$ and $g_{ti} =0$ (i.e. the shift is zero and the lapse is $N$) the evolution equations are 
\be\label{evolve}
\dot g_{ij} = -2N k_{ij}, \quad \dot k_{ij} = N({\cal R}_{ij}  +kk_{ij} -2k_{im}{k^m}_j-\Lambda g_{ij}) - D_i D_j N,
\ee
where a dot denotes $\partial /\partial t$.
These equations determine the second time derivative of $g_{ij}$ at every point in terms of the initial data. Taking a time derivative of these equations, one obtains an expression for the third time derivative in terms of initial data. Continuing in this way,  all time derivatives of   $g_{ij}$ at $t=0$ can be expressed in terms of the initial data. This is true at each point in space.  We can therefore take the limit as the point approaches the boundary at infinity. If we conformally rescale by an appropriate conformal factor,
 all time derivatives of the boundary metric at $t=0$ are completely fixed.

In general, these time derivatives can depend on the choice of conformal factor and choice of coordinates. One can discuss the corner conditions in terms of conformally invariant quantities \cite{Friedrich:1995vb,Enciso:2014lwa,Carranza:2018wkp}, however there are many situations of interest where a preferred conformal frame is picked out by symmetries. (An example is given in the next section.) 
The coordinate freedom in the metric can be dealt with as follows. In any conformal frame, one can introduce coordinates so the boundary metric takes the form $ds_\partial^2 = -dt^2 + q_{ij} dx^i dx^j$.  To ensure that $t$ agrees with an asymptotic time coordinate in the bulk, we choose the lapse to asymptotically approach $1/\Omega$ where $\Omega^2$ is the   conformal factor used to attach the conformal boundary. 
The net result is that all time derivatives of $q_{ij}$ at $t=0$ are fixed by the initial data.

\begin{figure}[ht]
    \centering
    \begin{subfigure}[b]{0.48\textwidth}
        \includegraphics[width=\textwidth]{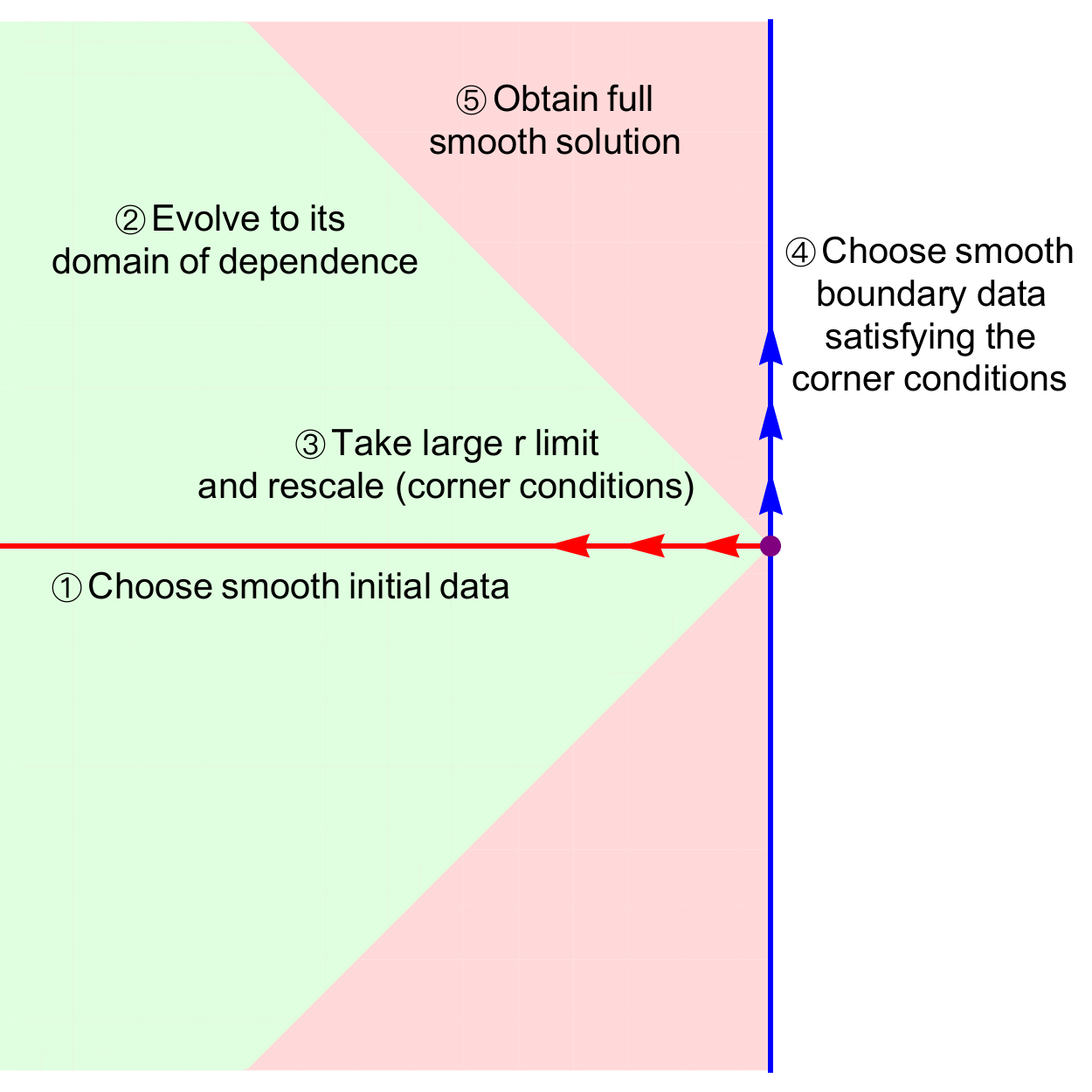}
        \caption{Initial boundary value problem.}
        \label{fig:recipe_corner}
    \end{subfigure}
    ~ 
    \begin{subfigure}[b]{0.48\textwidth}
        \includegraphics[width=\textwidth]{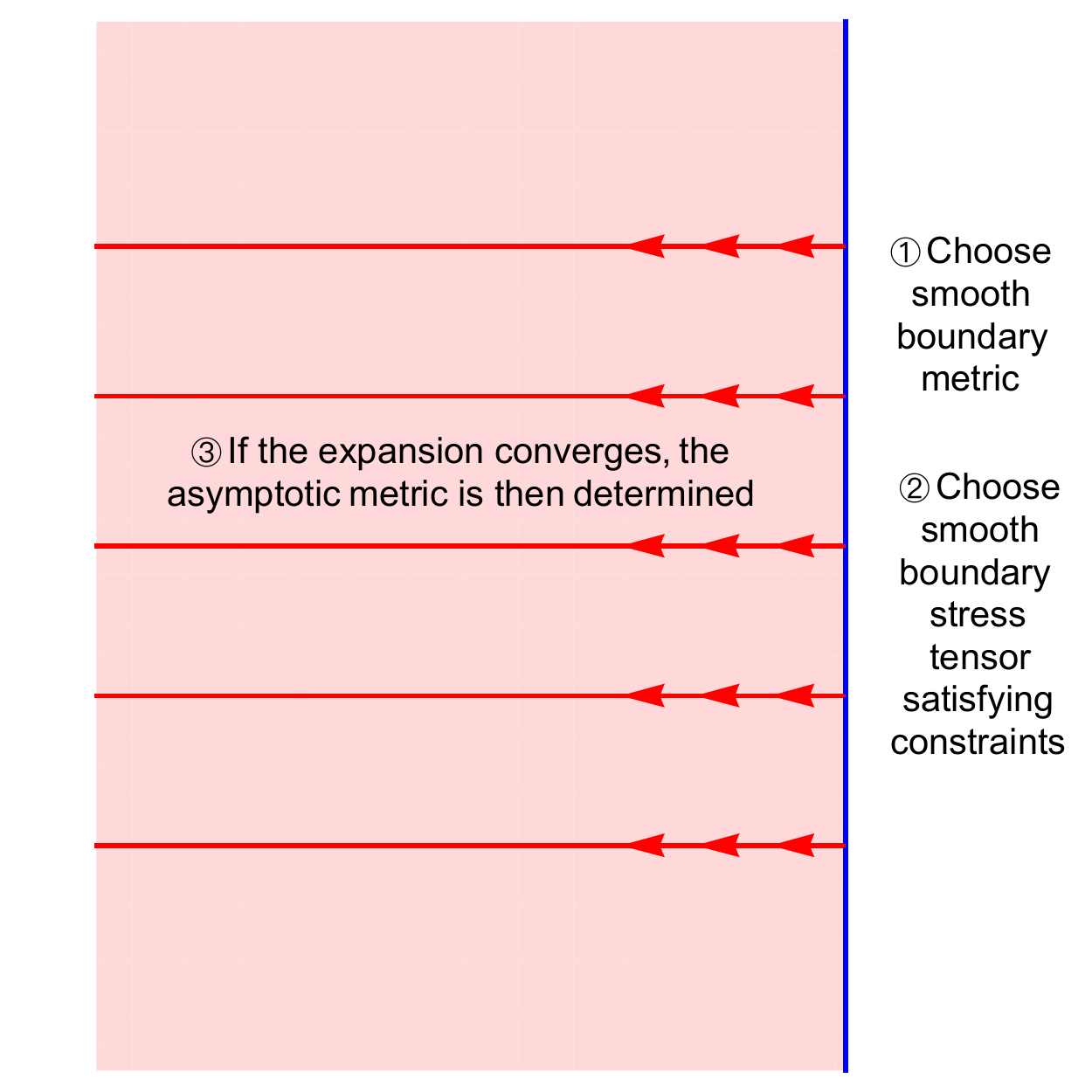}
        \caption{Pure boundary value problem.}
        \label{fig:recipe_fg}
    \end{subfigure}
    \caption{(a) Recipe for obtaining smooth bulk solutions given initial data and boundary data. (b) If the Fefferman-Graham expansion converges, one can obtain the asymptotic  bulk solution from the boundary metric and boundary stress tensor.}
\end{figure}

This does not contradict the usual picture of the domain of dependence of an asymptotically AdS spacelike surface, which shows that initial data at $t=0$ cannot determine the boundary metric at any $t>0$. There is still freedom to modify the boundary metric, but not in an arbitrary way. It must involve nonanalytic functions whose time derivatives at $t=0$ are all fixed. It is important to note that these corner conditions involve more than just a few leading corrections to the asymptotic metric. Higher order time derivatives are related to higher order spatial derivatives of the initial data, so all powers of $1/r$ are involved (see Fig.~\ref{fig:recipe_corner}).\footnote{ Even though the evolution equations (\ref{evolve}) involve two derivatives of $N$, once a conformal frame is chosen the remaining gauge freedom in choosing $N$ does not contribute to the time derivatives of the asymptotic metric.} 

One consequence of these corner conditions is that one cannot take initial data for an asymptotically AdS solution with a time dependent boundary metric, and ask how it would evolve with a static boundary metric. A simple example is the AdS analog of Witten's ``bubble of nothing" \cite{Witten:1981gj}. A double analytic continuation of the usual Schwarzschild AdS solution yields (setting the AdS radius to one)
\be\label{bubble}
ds^2 = \left (r^2 + 1 - \frac{M}{r} \right ) d\chi^2 + \left (r^2 + 1 - \frac{M}{r} \right )^{-1} dr^2 + r^2 ( -dt^2 + \cosh^2 t \ d\phi^2).
\ee
The $\chi$ coordinate is periodic and its period is chosen so that the circle it generates smoothly pinches off at $r=r_0$ where  $g_{\chi\chi}(r_0) = 0 $. The spacetime only exists for $r\ge r_0$.  Rescaling by $r^{-2}$ and taking the limit $r\to\infty$ one sees that the metric on the boundary at infinity is (conformal to)  a product of a circle and two dimensional de Sitter spacetime. Now consider the time symmetric initial data for this solution at $t=0$.  With the given boundary metric, this satisfies all the corner conditions.\footnote{This answers the question in \cite{Carranza:2018wkp} for a nontrivial example which satisfies all corner conditions.}
Suppose one wants to take the same initial data and evolve it with a static boundary metric. The corner conditions show this is impossible. We will return to this example in the last section and discuss how close one can come to realizing this solution.

A variety of results in general relativity involve initial data with asymptotic boundary conditions formulated in terms of powers of $1/r$. An important example is the positive energy theorem. Typically, the behavior of the first few terms are specified, but the subleading terms are arbitrary.  One might get the impression that the subleading terms have no significant effects. This is incorrect.  Another consequence of the corner conditions is that initial data sets which differ only in their subleading terms cannot be evolved with the same boundary metric. We will present an example of this in the next section.

Following the pioneering work of Chesler and Yaffe \cite{Chesler:2013lia}, much of the numerical work on time dependent holography has used initial data specified on an ingoing null surface which starts on a cross-section of the boundary. This initial data is not subject to an  infinite number of corner conditions. The difference can already be seen in the two dimensional wave equation on a half line $x <0$.  If $u=x-t$ and $v=x+t$ are null coordinates on Minkowski spacetime, and one is given smooth initial data $\phi(u,v=0) = f(u) $ for $u<0$ and smooth boundary data $\phi(x=0,t)$ for $t>0$, the only constraint is $f(0) = \phi(0,0)$. If this is satisfied,  there is always a smooth solution $\phi(u,v) = f(u)+g(v)$ in the region $u<0$ and $v>0$ where $g(\xi) = \phi(x=0,\xi) - f(-\xi)$. 

In holography, one often starts with pure AdS and adds time dependent sources on the boundary, or makes the boundary metric depend on time in the future.   If the sources (or metric) turn on at $t=0$ with any power of $t$, the boundary data will not be smooth, and the bulk solution will not remain smooth.  By causality, the lack of smoothness can only affect the region of the bulk to the causal future of the co-dimension two  $t=0$ surface on the boundary. However one can take  initial data on the ingoing null surface originating at $t=0$ and expect to obtain a smooth solution to its future (because the boundary data is smooth for $t > 0$ and we do not have an infinite number of corner conditions). This indicates that the lack of smoothness in this case is confined to the null surface.

Another way to avoid the corner conditions is to start with a static solution and modify the freely specifiable part of the initial data in a compact region. The corner conditions will then be satisfied with the same static boundary metric. For time symmetric initial data in AdS, one can take the conformal metric on the initial surface to be freely specified, and then solve the constraint (\ref{constraint}) for the conformal factor (with $k_{ij}=0$) \cite{Andersson:1992yk}.

So far, we have been discussing the initial value problem with boundary data. It is interesting to compare this to the purely ``boundary value problem", where only the boundary data are given. In this case, it is instructive to  use  the usual Fefferman-Graham expansion \cite{Fefferman:1985,Fefferman:2007rka}. If we write the asymptotic metric in the form
\be \label{FGmetric}
ds^2 = \frac{1}{z^2} [dz^2 + g_{\mu\nu}(z,x) dx^\mu dx^\nu],
\ee
and expand
\be\label{FGexp}
 g_{\mu\nu}(z,x) = \sum_{n=0}^\infty \gamma_{\mu\nu}^{(n)} z^n,
 \ee
 then it is known that in $3+1$ dimensions, $\gamma_{\mu\nu}^{(1)} =0$ and $\gamma_{\mu\nu}^{(2)}$ is fixed in terms of the curvature of the boundary metric $\gamma_{\mu\nu}^{(0)}$.  The next term, $\gamma_{\mu\nu}^{(3)}$, is not determined and represents the expectation value of the boundary stress tensor  $\langle T_{\mu\nu} \rangle= (3/16\pi G) \ \gamma_{\mu\nu}^{(3)}$ \cite{deHaro:2000vlm,Fischetti:2012rd}. If both the boundary metric and the boundary stress tensor are known for all $t$, then all terms in the expansion (\ref{FGexp}) are determined.  
 This series is known to converge in a neighborhood of the boundary for analytic data, but may not converge for general smooth data.  Fig.~\ref{fig:recipe_fg} illustrates this situation. (When the series converges, regularity in the interior is often used to fix $\gamma_{\mu\nu}^{(3)}$ in terms of $\gamma_{\mu\nu}^{(0)}$.) Here, corner conditions are not usually discussed simply because there is no corner anymore. However, at any point on the boundary, those relations between spatial derivatives and time derivatives of the metric are still valid. While the time derivatives were obtained from the spatial derivatives in the discussion of corner conditions, here it is the other way around: the boundary data and their time derivatives are used to determine the spatial derivatives of the metric at this point and thus the coefficients of the expansion in powers of $z$. In fact, we can think of this as a radial evolution system. 
 The fact that $\gamma_{\mu\nu}^{(3)}$ needs to be divergence free and traceless, is a consequence of the constraint equations for this system.
 
Before we discuss our example, we review an important aspect of the Fefferman-Graham expansion \cite{Fefferman:2007rka}. Starting with (\ref{FGmetric}) and (\ref{FGexp}), Fefferman and Graham set $\rho = z^2$ and write down a set of second order differential equations for $g_{\mu\nu}(\rho,x)$. When the boundary dimension $d$ is odd, they show that there is a power series solution in integer powers of $\rho$. This solution is uniquely determined by the boundary metric $\gamma_{\mu\nu}^{(0)}$.  This class of solutions all have zero energy since there is no $z^d$ term in the expansion.

 Solutions with nonzero energy can be obtained by adding a $\rho^{d/2}$ term to the expansion. But since the equations only involve $ \rho$ and $\partial/\partial \rho$, the next integer power of $\rho$ term that can be affected  is $ \rho^d$. This is because we need to combine two $\rho^{d/2}$ terms to affect an integer power, and the highest derivative terms in the equations take the form $g'' + g'^2 $ (where a prime denotes $\partial/\partial \rho$). This implies that in $3+1$ dimensions,  the $\gamma_{\mu\nu}^{(4)}$ term in the expansion (\ref{FGexp}) is determined entirely by $\gamma_{\mu\nu}^{(0)}$, and is not affected by the choice of $\gamma_{\mu\nu}^{(3)}$. Suppose we are given $\gamma_{\mu\nu}^{(0)}$ but not $\gamma_{\mu\nu}^{(3)}$, and we are asked to choose initial data on a spacelike surface. This argument shows that, in addition to the usual initial data constraints, the $\mathcal{O}(z)$, $\mathcal{O}(z^2)$ and $\mathcal{O}(z^4)$ terms are further constrained (in fact completely fixed). These are part of the corner conditions. This part also applies to data on an ingoing null surface.

\section{Example}

\label{sec:example}

In this section, we illustrate the corner conditions with an example. We start with a simple vacuum solution known as the AdS soliton \cite{Horowitz:1998ha}
\begin{align}
\label{eq:soliton}
ds^2 &= -r^2 dt^2 + \left(r^{2}-\frac{M}{r}\right) d\chi^2 + \left(r^{2}-\frac{M}{r}\right)^{-1}dr^2 + r^2 d\phi^2.
\end{align}
This metric is static and the boundary geometry is flat. To keep the metric smooth at $r=r_0 \equiv M^{1/3}$ where $g_{\chi\chi} =0$, $\chi$ must be periodic with period 
$\Delta \chi = 4\pi/3r_0$. We will choose to compactify $\phi$ as well with period $2\pi$. So the boundary is a static torus.
  
Now we ask what happens to the boundary metric if we take soliton initial data on a constant $t$ surface  and modify it.  It is clear from (\ref{bubble}) that if we replace the $r^{2}-{M}/{r}$ factors by $r^{2} +1-{M}/{r}$ the $\phi$ circle on the boundary becomes time dependent and expands exponentially.\footnote{We are fixing our conformal frame by demanding that the $\chi$ circle remains constant.}  Instead, we want to modify the soliton initial data by a higher order term in $1/r$ and ask what happens to the boundary.   To do this, it is convenient to consider a class of time-symmetric initial data
\begin{align}\label{initialdata}
ds^2|_{t=0} &=  \alpha(r) d\chi^2 + \frac{dr^2}{\alpha(r)\beta(r)} + r^2 d\phi^2,
\end{align}
which has to satisfy the constraint
\begin{align}
\label{eq:ode_beta}
(4 \alpha^\prime	+2 r \alpha^{\prime\prime}) \beta+(2 \alpha+r \alpha^\prime) \beta^\prime=12 r.
\end{align}
Since there is only one equation on two functions,  we can choose $\alpha(r)$ freely and determine $\beta(r)$ up to a constant. This form of the initial data can describe solutions where  the $\chi$ circle pinches off or one where it does not, depending on whether $\alpha(r)$ vanishes somewhere.  If it does pinch off at some positive $r$, then the interior can be made smooth by choosing the period of $\chi$ appropriately. 

To be specific, we will choose \be
\label{eq:initial_alpha}
\alpha(r) = r^{2}+\frac{a_1}{r}+\frac{a_2}{r^{2}}.
\ee
Solving (\ref{eq:ode_beta}) for $\beta$ yields the asymptotic solution
\be
\beta(r)=1+\frac{b_3}{r^3}+\frac{b_4}{r^{4}}+\cdots.
\ee
The coefficient $b_3$ is not fixed by the constraint equation and represents the free parameter in the solutions to this first order ODE. Since the leading term in $\alpha$ is $r^2$, if $b_3 \ne 0$, we have effectively made an $\mathcal{O}(1/r)$ change in the initial data. Since we want our modification to be strictly higher order, we will set $b_3=0$. $b_4$ and all higher order coefficients in $\beta$ are then fixed in terms of $a_1$ and $a_2$, which are the two parameters which label this class of solutions.
These two coefficients will be chosen so that  $\alpha(r_0)=0$ for some positive radius $r_0$, but otherwise they are not constrained. An explicit solution for $\beta$ can be written down in terms of an integral of exponentials of $\alpha, \alpha', $ and $\alpha''$, but it appears difficult to express the integral in terms of elementary functions. One can either evaluate the integral numerically, or simply solve (\ref{eq:ode_beta}) numerically.

It is clear that under evolution, the boundary metric cannot stay static. In terms of Fefferman-Graham coordinates, $z = (1/r) [1-a_1/6r^3 +\mathcal{O}(r^{-4})]$, we have made an $\mathcal{O}(z^4)$ change in the initial data. As reviewed at the end of the previous section, if the boundary metric remained flat, both the $\mathcal{O}(z^2)$ and $\mathcal{O}(z^4)$ terms in the expansion of the solution would have to vanish. We now compute the time dependence induced by this simple change to the initial data.

To describe evolution off the surface, we promote the coefficients to functions of time and add a $g_{tt}$ component to the metric. In a neighborhood of this initial spatial surface ($t=0$), we write
\be\label{evolution}
ds^2 = G(r,t) dt^2 + A(r,t) d\chi^2 + \frac{dr^2}{A(r,t)B(r,t)} + r^2 F(t) d\phi^2.
\ee
Note that we have used diffeomorphism freedom in $r$ and $t$ to set $g_{rt}=0$ and $g_{\phi\phi}=r^2 F(t)$. Since we are interested in the asymptotic form of the solution, we expand
\begin{align}
A(r,t)&=r^{2}+A_0(t)+\frac{A_1(t)}{r}+\frac{A_2(t)}{r^{2}}+\cdots,\\
B(r,t)&=B_0(t)+\frac{B_1(t)}{r}+\frac{B_2(t)}{r^{2}}+\frac{B_3(t)}{r^{3}}+\frac{B_4(t)}{r^{4}}+\cdots,\\
G(r,t)&=-r^{2}+G_0(t)+\frac{G_1(t)}{r}+\frac{G_2(t)}{r^{2}}+\cdots,\\
F(0)&=1, \quad \; A(r,0) = \alpha(r), \quad \; B(r,0) = \beta(r),
\end{align}
where the leading order term in $G(r,t)$ is fixed using residual diffeomorphism freedom in $t$. The general four-dimensional  metric with $U(1)^2$ symmetry can be described by three functions of $(r,t)$. It may seem odd that our metric has three functions of $(r,t)$, and in addition a function $F(t)$. However, $F(t)$ can be viewed as taking the place of the time dependence in the leading order term in $A(r,t)$. We have required this to be just $r^2$ so we can use $1/r^2$ as our conformal rescaling to a boundary metric with constant size $\chi$ circles.  The boundary metric is simply
\be\label{bdymetric}
ds^2 = -dt^2 + d\chi^2 + F(t) d\phi^2.
\ee

\begin{sidewaystable}[htbp] 
  \footnotesize
  \begin{center}
  \caption{Time derivatives of the solution (\ref{evolution}) at $t=0$. The number of time derivatives is listed at the top of each column. }
  \label{tab:table1}
  \begin{tabular}{|c|>{\centering}p{2.7cm}|>{\centering}p{2.7cm}|>{\centering}p{2.7cm}|>{\centering}p{2.7cm}|>{\centering}p{2.7cm}|>{\centering}p{2.7cm}|>{\centering}p{2.7cm}|}
    \hline
    & 0 & 2 & 4 & 6 & 8 & 10 & 12 \tabularnewline \hline
    \hline
    $F$ &1&0&8 $a_2$ &0&4224 $a_2^2$&15840 $a_1^2 a_2$& 21997568 $a_2^3$\tabularnewline 
    \hline \hline
    $G_0$ & 0 & 0 & 0 & 0 & 0 & 0&  \tabularnewline \hline
    $G_1$ & 0 & 0 &  $-$50 $a_1 a_2$ & 0 & $-$172228 $a_1 a_2^2$ && \tabularnewline  \hline
    $G_2$ & 0 & 0 & 0 & 0 & 0 &  &  \tabularnewline \hline
    $G_3$ & 0 & 5 $a_1 a_2$ & 0 & 93114 $a_1 a_2^2$/5 &&& \tabularnewline \hline
    $G_4$ & 0 & 52 $a_2^2$/3 & 135 $a_1^2 a_2$ & 271168 $a_2^3$ /3 &&& \tabularnewline \hline
    $G_5$ & 15 $a_1 a_2$/28 & 0  & 123321 $a_1 a_2^2$/70 &&&& \tabularnewline \hline
    $G_6$ & 4 $a_2^2$/5 & 13 $a_1^2 a_2$/16 & 53456 $a_2^3$/15 &&&& \tabularnewline \hline
    $G_7$ & 0 & 4601 $a_1 a_2^2$/252 & & &&& \tabularnewline \hline
    $G_8$ & $-$21 $a_1^2 a_2$/40 & 868 $a_2^3$/25 & & &&& \tabularnewline \hline
    $G_9$ &  $-$2393 $a_1 a_2^2$/1760 & & & &&& \tabularnewline \hline
    $G_{10}$ & $-$8 $a_2^3$/9 &  & & &&& \tabularnewline \hline
    \hline
    $A_0$ &  0&  4 $a_2$&0 & 1312 $a_2^2$	&7920 $a_1^2 a_2	$&6544384 $a_2^3$& \tabularnewline \hline
    $A_1$ &  $a_1$	&0&	100 $a_1 a_2$	&0	&365456 $a_1 a_2^2$ && \tabularnewline \hline
    $A_2$ &  $a_2$	&0	&344 $a_2^2$	&1980 $a_1^2 a_2$	&1717504 $a_2^3$&& \tabularnewline \hline
    $A_3$ & 0	&10 $a_1 a_2$&	0&	187728 $a_1 a_2^2$/5 &&& \tabularnewline \hline
    $A_4$ & 0&	16 $a_2^2$	& 85 $a_1^2 a_2$	&262976 $a_2^3$/3 &&& \tabularnewline \hline
    $A_5$ & 0	&0&	31632 $a_1 a_2^2$/35 & &&& \tabularnewline \hline
    $A_6$ &  0	&$-$9 $a_1^2 a_2$/4 &20864 $a_2^3$/15 & &&& \tabularnewline \hline
    $A_7$ &  0	&-117 $a_1 a_2^2/35$ & & &&& \tabularnewline \hline
    $A_8$ & 0 &0  & & &&& \tabularnewline \hline
    $A_9$ & 0 &  & & &&& \tabularnewline \hline
    $A_{10}$ & 0 &  & & &&& \tabularnewline \hline
    \hline
    $B_0$ &1  &0  &0 &0 &0&0&0 \tabularnewline \hline
    $B_1$ & 0 &0  &0 &0 &0&0&\tabularnewline \hline
    $B_2$ &  0&  0& 0& 0&0&0& \tabularnewline \hline
    $B_3$ &0	&0	&50 $a_1 a_2$&	0&	172228 $a_1 a_2^2$&& \tabularnewline \hline
    $B_4$&  $a_2$ &0&	344 $a_2^2$ &1980 $a_1^2 a_2$&	1717504 $a_2^3$&& \tabularnewline \hline
    $B_5$&  0&	15 $a_1 a_2$&	0&	275343 $a_1 a_2^2/5$ &&& \tabularnewline \hline
    $B_6$ &  0&	32 $a_2^2$ &	315 $a_1^2 a_2/2$ & 511552 $a_2^3/3$ &&& \tabularnewline \hline
    $B_7$ &  $-$ $a_1 a_2 /4$ &	0&	31023 $a_1 a_2^2 /14$ & &&& \tabularnewline \hline
    $B_8$ & $a_2^2 /5 $&	$-$21 $a_1^2 a_2 /2$ &	23248 $a_2^3 /5$ & &&& \tabularnewline \hline
    $B_9$ & 0	& $-$19 $a_1 a_2^2 /20 $& & &&& \tabularnewline \hline
    $B_{10}$ &   $a_1^2 a_2 /16$	&192 $a_2^3 /5$  & & &&& \tabularnewline \hline
    $B_{11}$ &  $-$13 $a_1 a_2^2 /160$ &  & & &&& \tabularnewline \hline
    $B_{12}$ & $a_2^3 /45$ &  & & &&& \tabularnewline
    \hline
  \end{tabular}
  \end{center}
\end{sidewaystable}

The vacuum Einstein equations can then be used to determine the time derivatives of the metric, including the boundary metric component $F(t)$. To do this, we expand the Einstein equations in powers of $1/r$ and require it to vanish at each order. Then we take time derivatives of Einstein equations and do the same. Taking more time derivatives of the Einstein equations allows computation of higher time derivatives of the metric components. See Table \ref{tab:table1} for the results up to 12 time derivatives. Since our initial data is time symmetric, an odd number of time derivatives of any quantity vanishes, so only an even number of time derivatives appear in the table. Note that only the top left half of each section of the table are filled. This is related to how the values are calculated in practice, as $A(r)$ is expanded only to a finite order and the $(n+2)$-th time derivatives of the $\mathcal{O}(1/r^m)$ coefficients are related to the $n$-th time derivatives of the $\mathcal{O}(1/r^{m+2})$ coefficients in general. More entries can of course be calculated if we expand $A(r,t)$ to higher powers of $1/r$ to compute higher-power coefficients of Einstein equations.   As expected, all time derivatives vanish if $a_2=0$.

To the order shown in the table,  derivatives of the function $F(t)$ at $t=0$ are completely determined by the initial data, and  this will continue to higher orders. This illustrates the corner conditions discussed in the previous section. Note that the derivatives of $F(t)$  grow very rapidly. This shows that $F(t)$, if analytic, would grow much faster than $e^t$. The Taylor series begins
\be\label{expan}
F(t) =1+ \frac{1}{3} a_2 t^4 + \frac{11}{105} a_2^2 t^8 + \frac{11}{2520} a_1^2 a_2 t^{10} + \frac{21482}{467775} a_2^3 t^{12} + \cdots.
\ee
If we view $F(t)$ as a function of $\tau = t^2$, one finds that derivatives with respect to $\tau$ are still growing, suggesting that $F(t)$ is growing faster than $e^{t^2}$. It is surprising that a subleading change in the initial data causes such a dramatic change in the boundary geometry. The fact that the time dependence of $F(t)$ begins with a $t^4$ term is expected, since a $t^2$ term would produce nonzero spacetime curvature at $t=0$. This would require a nonzero $z^2$ term in a Fefferman-Graham expansion, and our initial data does not contain such a term. Another interesting feature of this table is that to the order calculated, it appears that $A_2(t) = B_4(t)$ and $G_1(t) = -B_3(t)$.

A constant rescaling of the boundary metric (\ref{bdymetric}) will rescale the proper time and hence all time derivatives. But this also rescales the proper length of the $\chi$ circle. So the expansion  (\ref{expan}) is valid in the conformal frame in which the length of the $\chi$ circle is given by regularity in the interior of the bulk solution with the given $a_1, a_2$. This is part of a general scaling symmetry of our solution (\ref{evolution}). The metric is invariant under 
\be\label{scaling}
r= \lambda \tilde r, \ \ (t,\chi,\phi) = (\tilde t,\tilde \chi, \tilde\phi) / \lambda,\ \  (A,G) = \lambda^2 (\tilde A,  \tilde G),\ \  (B,F) = (\tilde B,\tilde F).
\ee
If we define the dimension of a quantity to be the power of $\lambda$ that it acquires under this transformation, then $a_1$ has dimension three and $a_2$ has dimension four.  Noting that each time derivative adds one to the dimension, one can check that the entries in Table 1 all have the correct dimension.

\begin{figure}[ht]
\begin{center}
\includegraphics[width=.7\textwidth]{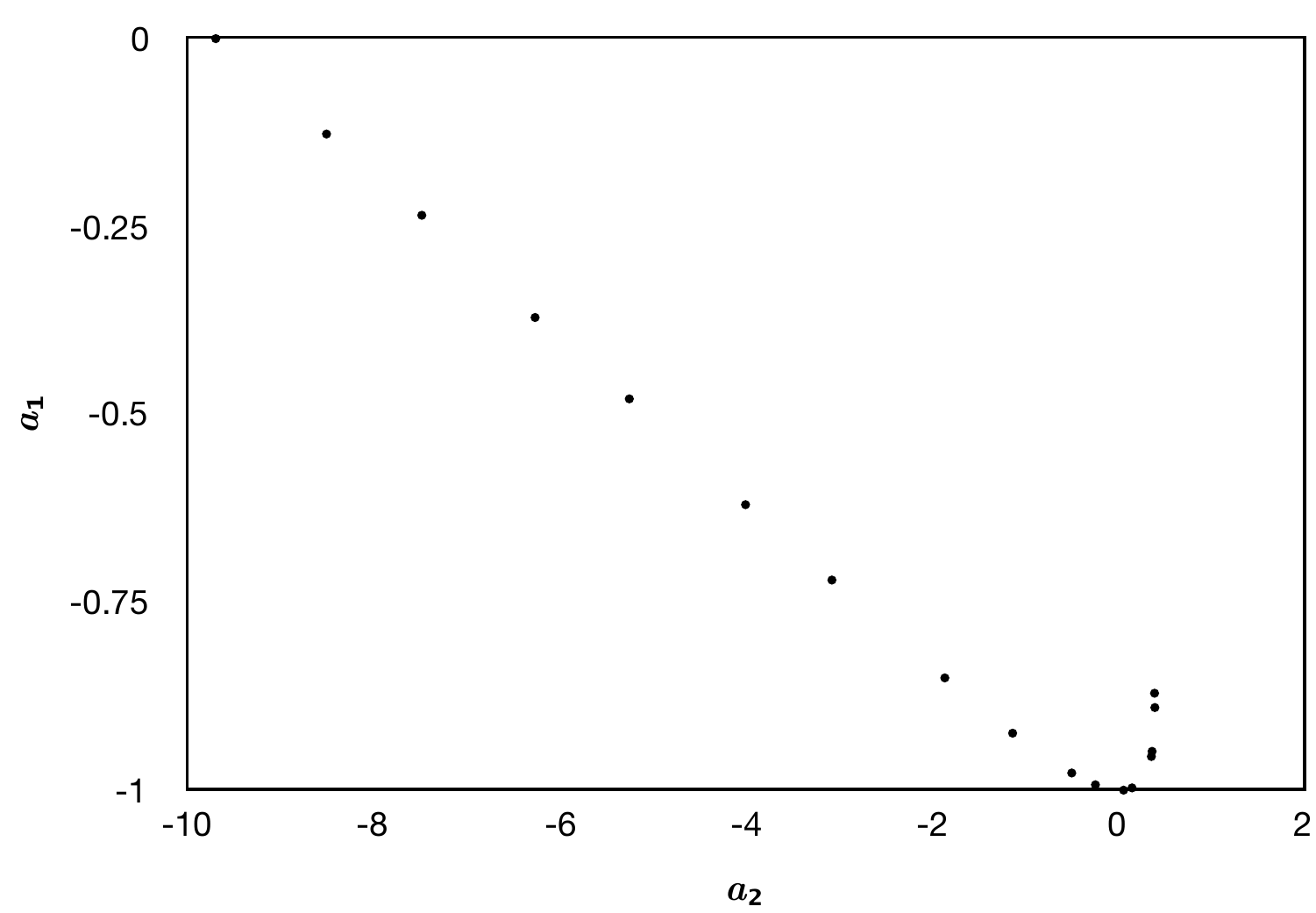}
 \caption{\label{fig:fixed_s_in_a1a2} Numerically computed curve in the parameter space of $a_1$ and $a_2$ that corresponds to fixing the periodicity of $\chi$ to be $4\pi/3$, which agrees with the AdS soliton with $M=1$. Corresponding curves for other periodicities can be easily obtained using the scaling symmetry (\ref{scaling}). Since the energy is proportional to $a_1$, this shows that it is minimized at the AdS soliton. }
 \end{center}
\end{figure}

It is also interesting to investigate the energy of these solutions. At first sight, it might seem that the initial data (\ref{initialdata})  with $\alpha$ given by (\ref{eq:initial_alpha}) and $b_3 = 0$ would have the same energy as the AdS soliton (with $a_1=-M$), since we have not changed the $O(z^3)$ term in the Fefferman-Graham expansion, so they have the same stress tensor. However, the stress tensor needs to be integrated over the boundary volume to give the energy. Recall that the periodicity of $\chi$ is chosen to make the interior smooth. In fact, it needs to be
\begin{align}
\Delta \chi = \frac{4\pi}{\alpha^\prime(r_0)\sqrt{\beta(r_0)}}.
\end{align}
Therefore,  the energy should be compared to an AdS soliton that has the same size circles on the boundary. Imposing this condition requires that $a_1$ increase with $|a_2|$ (see Fig.~\ref{fig:fixed_s_in_a1a2}).  Since $a_1$ is the energy density and the volume of space is now the same, this shows that the new solutions always have greater energy than the soliton. This is expected  since it was conjectured in \cite{Horowitz:1998ha} that the soliton minimizes the energy with these boundary conditions. In fact, for initial data with $U(1)^2$ symmetry, like the cases we are considering, this conjecture has recently been proven  \cite{Barzegar:2019vaj}. Note that nothing unusual happens in 
Fig.~\ref{fig:fixed_s_in_a1a2} when $a_1 = 0$. At this point, the stress tensor vanishes, showing this solution has the same energy as pure AdS. However, the AdS soliton has less energy than pure AdS, so the solution with $a_1 = 0$ and $a_2<0$ still represents a nontrivial excitation above the soliton ground state.

\section{Discussion}

In the introduction we mentioned that one cannot take initial data for the solution (\ref{bubble}) and evolve it with a static boundary metric.  But there is an interesting physics question related to this. First note that the interpretation of (\ref{bubble}) as a ``bubble of nothing" in AdS is ambiguous. Asymptotically the two circles parameterized by $\chi$ and $\phi$ are on equal footing. If one views the $(r,\phi)$ plane as a base, and the $\chi$ circle as a fiber over each point, then it looks like there is a hole in the $(r,\phi)$ plane, which expands out. This is the idea behind a ``bubble of nothing".  However, one can also choose to view the $(r,\chi)$ plane as a base, and the $\phi$ circle as a fiber, in which case there is no hole. The invariant statement is that there is a minimal circle at $ r=r_0$ and constant $t$ which expands forever.\footnote{In asymptotically flat spacetime, there is no ambiguity since the $\chi$ circle approaches a constant size at infinity while the $\phi$ circle grows. So only the $(r,\phi)$ plane is asymptotically flat.}

 It is natural to ask if this is only possible because one of the two circles on the boundary is also becoming large. To satisfy the corner conditions the boundary metric must look like $S^1\times dS_2$ near $t=0$, but away from $t=0$ we can stop the second circle from expanding and keep the boundary metric static to the future. (This change must be made in a smooth but nonanalytic way.) What will the evolution look like? The minimal circle will start to expand since inside the domain of dependence, the evolution is independent of the boundary conditions at infinity. But when it knows that the boundary is static, will it still become arbitrarily large, or will it reach a maximum and contract? It would be interesting to study this solution changing the boundary conditions a short time $\epsilon$ after $t=0$, and taking the limit $\epsilon \to  0$.

If one satisfies only the first $p$ corner conditions, the solution cannot be any smoother than $C^p$. But it might be much less regular. It was shown in \cite{Enciso:2014lwa} that in $3+1$ dimensions, the solution will be at least $C^q$ where $q < (p-15)/2$. The authors stress that they have not tried to maximize $q$, but no better result seems to exist in the literature.

What are the consequences of this loss of regularity for holography?\footnote{We thank Henry Maxfield and Don Marolf for discussions on this question.} The best understood examples of holography relate string theory with asymptotically AdS boundary conditions to a dual field theory. Physicists are often happy to assume all fields are smooth ($C^\infty$) and not worry about the difference between, e.g., $C^7$ and $C^{11}$ solutions. But the existence of the corner conditions means that there are situations of interest with smooth initial data and smooth boundary data, where the resulting solution will not be smooth. Note that this is different from the case discussed in Sec.~1 where the boundary data itself was not smooth. In general relativity, as long as the solution is $C^2$ (so the curvature is well defined), this does not seem to matter. However string theory is different. The usual Einstein equations arise as just the leading term in the classical equations of motion. There are higher order $\alpha'$ corrections that involve higher powers and derivatives of the curvature. If the bulk solution is smooth with curvature below the string scale, these higher order corrections can be taken into account with small perturbations to the original  solution. However, if the solution is not smooth,  at some point these higher order corrections may diverge, indicating that the leading solution is not close to an exact classical string theory solution everywhere. 

By causality, the violation of the corner conditions cannot affect the solution inside the domain of dependence. With smooth initial data, this will remain smooth, and close to an exact string solution. The key question is what happens outside this domain of dependence.  It is likely that the lack of smoothness will be concentrated on an ingoing null shock wave that originates where the corner conditions are violated. If so, violating the corner conditions would just produce a milder version of the gravitational shock waves \cite{Sfetsos:1994xa} that have been extensively studied in the context of holography (see, e.g., \cite{Shenker:2013pqa,Shenker:2013yza,Stanford:2014jda}). In some cases one can argue that all $\alpha'$ corrections to the shock wave vanish since the curvature is null \cite{Horowitz:1999gf}.
In this case the solution everywhere could  remain close to an exact string solution. 
However if the lack of smoothness propagates inside the future of the corner, then there will likely be large stringy corrections, and the leading order solution given by general relativity cannot be trusted there. It is clearly of interest to settle this question.

What about the dual field theory? Quantum field theory states can be defined at one moment of time, i.e., on one spacelike surface. This is clear in a Schr\"{o}dinger representation and realized in path integral definitions of states. In holography, the bulk state should correspond to a state in the dual QFT. One  expects that given a quantum state at $t=0$, one can evolve with any time dependent metric one chooses. One also expects that if an initial state is dual to semiclassical initial data for the bulk geometry, the evolved  state will continue to be dual to the semiclassical bulk. This is clearly in tension with the corner conditions. In light of this, it is important to remember 
 that many operators of interest cannot be defined unless one is given the spacetime in a neighborhood of the $t=0$ surface. For example,  in four dimensions, the expectation value of the trace of the stress energy operator involves the square of the curvature and two derivatives of the curvature. So the stress tensor cannot be defined just knowing the initial metric on a spacelike surface. One requires knowledge of at least four derivatives of the metric off the surface. This is relevant since the Hamiltonian which evolves the state is constructed from the stress tensor. The corner conditions suggest that there is a connection between the state and these time derivatives, at least for semiclassical states. Clarifying this connection may lead to a deeper understanding of holography.

\section*{Acknowledgements}

We thank D. Marolf, H. Maxfield, D. Mateos, J. Santos, and C. Warnick for discussions. This work was supported in part by NSF grant PHY1801805. 
\bibliographystyle{JHEP}
\bibliography{library}

\end{document}